\documentclass[twocolumn,prl,aps]{revtex4-2}
\usepackage{floatflt}
\usepackage{psfrag}
\usepackage{amsmath}   
\usepackage{amssymb}
\usepackage{amsfonts}
\usepackage{mathrsfs}
\usepackage{graphicx}

\usepackage[dvipsnames]{xcolor}
\usepackage[colorlinks=true, citecolor=Cerulean,linkcolor=blue, urlcolor = Cerulean]{hyperref}

\begin{document}

\title
{On the Nature of the Fundamental Plasma Excitation in a Plasmonic Crystal}

\author{A.~R.~Khisameeva$^{a}$, A.~Shuvaev$^{a}$, A.~A.~Zabolotnykh$^{b}$, A.~S.~Astrakhantseva$^{c}$, D.~A.~Khudaiberdiev$^{a}$, A.~Pimenov$^{a}$, I.~V.~Kukushkin$^{c}$, V.~M.~Muravev$^{c}$ \footnote{Corresponding E-mail: muravev@issp.ac.ru}}
\affiliation{$^a$ Institute of Solid State Physics, Vienna University of Technology, 1040 Vienna, Austria \\
$^b$ Kotelnikov Institute of Radio-engineering and Electronics of the RAS, Mokhovaya 11-7, Moscow 125009, Russia\\
$^c$ Institute of Solid State Physics, RAS, Chernogolovka, 142432 Russia
}

\date{\today}

\begin{abstract}
We report on the experimental study of the spectrum of plasma excitations in a plasmonic crystal fabricated from the two-dimensional electron system in an AlGaAs/GaAs semiconductor heterostructure. We perform a comprehensive research on the mode frequency and relaxation as a function of the gate width across different plasmonic crystal periods. Importantly, we develop an analytical approach that accurately describes the behavior of plasma excitations in plasmonic crystals, providing new insights into the fundamental physics of plasmonic systems.
\end{abstract}

\maketitle

A plasmonic crystal represents a two-dimensional electron system (2DES) with a periodic gate deposited on the crystal surface over the 2DES. Pioneering experiments on plasma waves in semiconductor two-dimensional systems have been conducted using plasmonic crystals more than 40 years ago~\cite{Allen:1977, Theis:1977, Heitmann:1982}. Since then, this system has become a platform for discovering numerous new physical phenomena. These include observation of minigaps in the plasmon spectrum~\cite{Heitmann:1984, Heitmann:1986}, 2D plasmons in layered materials~\cite{Wang:2011, Basov:2018}, plasmonic crystal Tamm states~\cite{Shaner:2012, Shaner:2013}, and various plasmonic crystal phases~\cite{Shur:2010, Knap:2023, Knap:2024}. Moreover, most of the currently developed plasmonic devices rely on the concept of plasmonic crystals. Examples include terahertz (THz) radiation detectors~\cite{Allen:2002, Knap:2009, Lusakowski:2014}, THz oscillators and amplifiers~\cite{Tsui:1980, Mikhailov:1998, Popov:2020}, and THz phase shifters~\cite{Muravev:2022, Muravev:2023}. 

The currently prevailing theoretical model suggests that an incident electromagnetic wave excites two types of plasma waves in the plasmonic crystal. When the distance from the gate to the 2DES is relatively small, this model is analogous to the tight-binding limit of the Kronig-Penney model, which describes the particle behavior in a periodic potential~\cite{Aizin:2012, Popov:2015, Kachorovskii:2012, Svintsov:2019, Kachorovskii:2024}. The plasma wave in the ungated region, which is excited in the gap between the gates, is described by the spectrum~\cite{Stern:1967}: 
\begin{equation}
    \omega_u(q) = \sqrt{\frac{n_s e^2}{2 m^{\ast} \varepsilon(q) \varepsilon_0} \, q_u},
\label{plasmon}
\end{equation} 
where $n_s$ is the 2D electron density, $m^{\ast}$ is the effective electron mass, $q_u$ is the unscreened plasmon wave vector, $\varepsilon(q)$ is the effective dielectric permittivity of the medium surrounding the 2DES. In the region directly underneath the gate, the plasma wave is the screened plasmon with the linear spectrum~\cite{Chaplik:1972}
\begin{equation}
\omega_g(q) = \sqrt{\frac{n_s e^2 h}{m^{\ast} \varepsilon \varepsilon_0}} \, q_g \qquad (q_g h \ll 1).
\label{scr_plasmon}
\end{equation} 
Here $h$ is the distance between the gate and the 2DES, $q_g$ is the screened plasmon wave vector, and $\varepsilon$ is the dielectric permittivity of the semiconductor crystal, specifically $\varepsilon_{\rm GaAs} = 12.8$. Then, the existing theoretical models make several assumptions. 
The plasmon dispersion relation for the system is obtained by (i) plane-wave
expansion of the plasma wave electric potential in each region, (ii) matching of the potential and electric current at the gated-ungated boundaries, and (iii) imposing Bloch periodicity
conditions for the fields in the neighboring gratings. These assumptions lead to Kronig-Penney-type dispersion of the optically active modes~\cite{Aizin:2012, Popov:2015, Kachorovskii:2012, Svintsov:2019, Kachorovskii:2024}:
\begin{equation}
\label{KP}
    q_u \sin\left(\frac{q_g l_g}{2} \right) \cos\left(\frac{q_u l_u}{2} \right) + q_g \cos\left(\frac{q_g l_g}{2} \right) \sin\left(\frac{q_u l_u}{2} \right) = 0,
\end{equation}
where $l_g$ is the gate strips width, and $l_u$ is the width of the ungated 2DES area between the strips.

In the present work, we investigate the terahertz (THz) spectrum and relaxation of plasma excitations in a plasmonic crystal fabricated from an AlGaAs/GaAs heterostructure. Our experimental data reveal that the plasmonic crystal supports a series of resonant plasmon modes. We directly trace how the fundamental plasmon mode evolves from an unscreened plasmon, Eq.\,(\ref{plasmon}), for narrow gates to a screened plasmon, Eq.\,(\ref{scr_plasmon}), for wider gates.  Importantly, we have developed an analytical model that accurately describes the dispersion and relaxation of plasma waves propagating through the plasmonic crystal. The theory convincingly establishes that plasma excitations excited in the plasmonic crystal possess a delocalized superlattice nature. Our results hold promise for engineering of THz plasmonic devices.

\begin{figure}[t!]
    \centering
    \includegraphics[width=\linewidth]{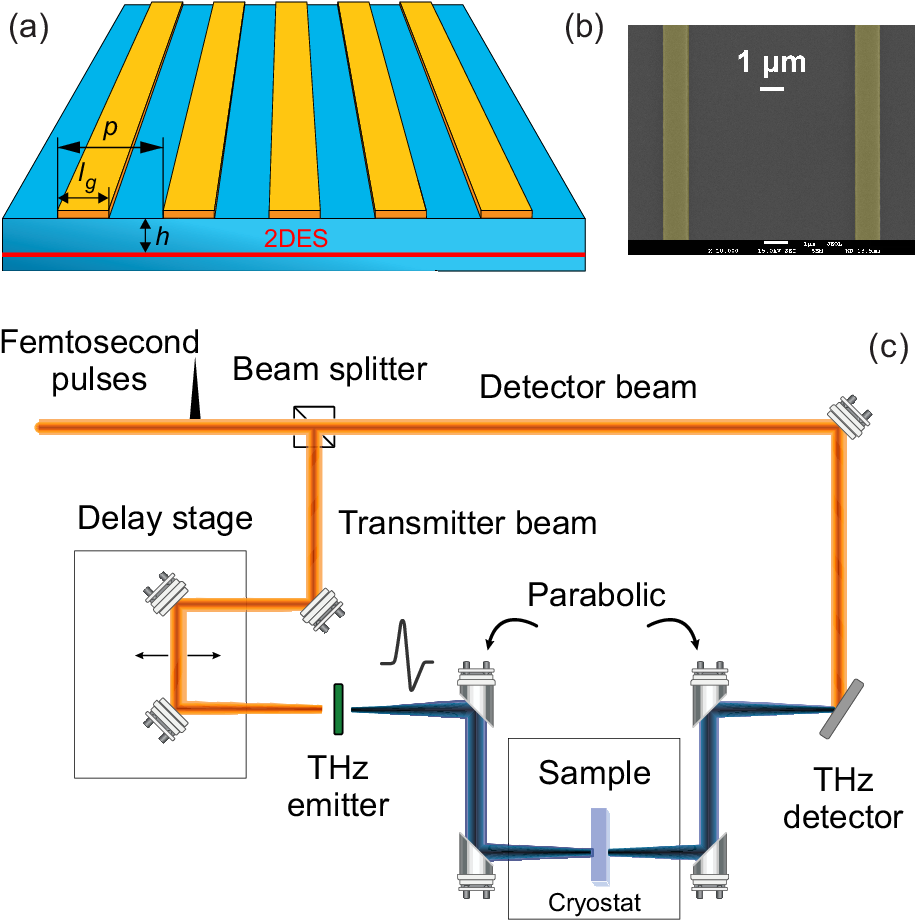}
    \caption{(a) Sketch of the plasmonic crystal. (b) Scanning electron-beam microscope image of the grid-shaped gate. (c) Schematic diagram of the experimental TDS setup, highlighting the key elements of the apparatus.}
    \label{1}
\end{figure}

In our study, we have explored the samples which were fabricated from the Al$_{0.24}$Ga$_{0.76}$As/GaAs/Al$_{0.24}$Ga$_{0.76}$As heterostructure grown by the molecular beam epitaxy with a GaAs quantum well of $20$~nm thickness situated at $210$~nm beneath the crystal surface. Throughout the paper, we used the distance from the center of the quantum well to the crystal surface, $h=220$~nm. The density of two-dimensional electrons in the quantum well was $n_{s}=1.47\times 10^{12}~\text{cm}^{-2}$ with  mobility $\mu = 101 000~\text{cm}^2/\text{V$\cdot$s}$, derived from the high-frequency data. A Cr($24$~nm)/Au($200$~nm) grid-shaped gate is evaporated on the top surface of the sample (Fig.~\ref{1}(a,b)). The samples made using E-beam lithography have 
Cr($5$~nm)/Au($95$~nm). The grid strips have a width from $l_g=0.8$~\textmu{}m up to $10$~\textmu{}m. We fabricated two sets of samples with two fixed grid periods of $p = 8$~\textmu{}m and $12$~\textmu{}m. A detailed table of all measured samples is listed in the Supplemental Material~\cite{Supplemental}. The experiments were conducted using the time-domain spectroscopy setup, which covers a frequency range from $0.15$ to $2$~THz, see Fig. 1(c). 
The spectrometer is equipped with a continuous-flow liquid helium cryostat with transparent windows. All measurements are performed at the base sample temperature of $T=5$~K. To minimize noise from water absorption lines in the air, the THz beam path is enclosed in an evacuated chamber. Spectroscopy measurements are carried out using a $6$~mm aperture positioned close to the sample to ensure that the electromagnetic radiation passes only through the grating-gate active region of the plasmonic crystal. The polarization of the electromagnetic radiation is directed perpendicular to the grids.


\begin{figure}[!t]
    \centering
    \includegraphics[width=0.95\linewidth]{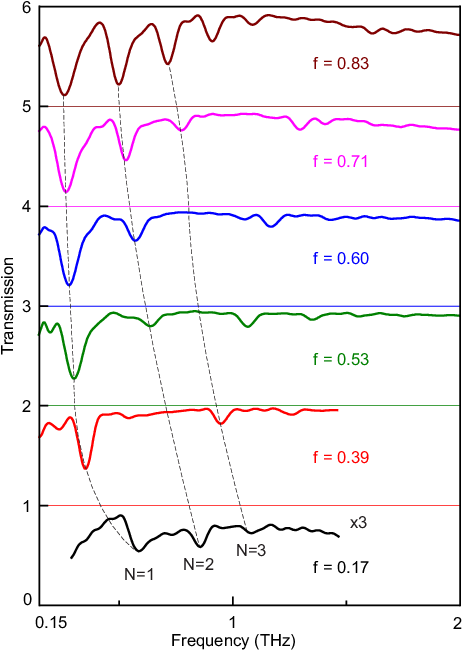}
    \caption{Transmission spectra measured for the plasmonic crystal with different grid filling factors $f=0.17, 0.39, 0.53, 0.60, 0.71$ and $0.83$. These filling factors correspond to gate widths of $l_g = 2.0, 4.7, 6.3, 7.2, 8.5$, and $9.9$~\textmu{}m, respectively. The grid period is kept fixed at $p = 12$~\textmu{}m. The dotted lines is a guide to the eye, connecting the frequency positions of the fundamental, $N=1$, and harmonic, $N=2,3$ plasma resonances. The polarization of the electromagnetic wave is perpendicular to the grids.}
    \label{2}
\end{figure}

\begin{figure*}[!t]
    \centering
    \includegraphics[width=0.99\linewidth]{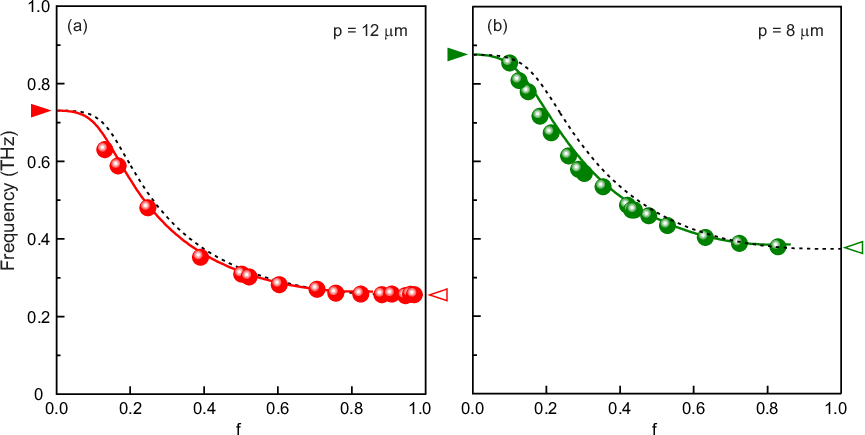}
    \caption{The frequency of the fundamental plasma mode in the plasmonic crystal as a function of the filling factor $f = l_g/p$. The period of the plasmonic crystal is fixed at $p = 12$~\textmu{}m (a) and $p = 8$~\textmu{}m (b). Solid lines show the result of our theoretical model~(\ref{eq:theor_freq}). Dashed line represents the prediction from the Kronig-Penney plasmonic model Eq.~(\ref{KP})~\cite{Supplemental}. The filled arrows point to the frequency of the unscreened plasmon from Eq.~(\ref{plasmon}) with wave vector $q=2 \pi/p$. The empty arrows point to the frequency of the screened plasmon from Eq.~(\ref{scr_plasmon}) with the same wave vector $q=2 \pi/p$.}
    \label{3}
\end{figure*}

Figure~\ref{2} shows the transmission through a series of samples with varying gate widths, 
keeping the grid period constant at $p = 12$~\textmu{}m. We refer to the Supplemental Material section for the analogous set of curves recorded for the plasmonic crystal with a period of $p = 8$~\textmu{}m~\cite{Supplemental}. For clarity, the curves are shifted along the vertical axis. Each sample exhibits several plasma resonances in the transmission spectrum. The dotted line marks the frequency position of the fundamental plasma resonance, $N=1$, as well as harmonic $N=2,3$ modes. As the gate width decreases, the frequency of the fundamental plasma mode increases. Figure~\ref{3} summarizes the frequency positions of the fundamental plasma mode versus the grid filling factor $f = l_g/p$ measured for the plasmonic crystal with $p = 12$~\textmu{}m (a) and $8$~\textmu{}m (b). The dashed black lines in Fig.~\ref{3} depict the frequency of the plasmon modes predicted from the Kronig-Penney model Eq.~(\ref{KP})~\cite{Supplemental}. In computations, we take $m^{\ast} = (0.076 \pm 0.001) \, m_0$ due to the nonparabolicity of the electron conduction band occurring at such high 2D electron densities of $n_{s}=1.47\times 10^{12}~\text{cm}^{-2}$~\cite{Francisco:1988, Ekenberg:1989, Muravev:PRA2023}. The results from model~(\ref{KP}) show a reasonable agreement with the experimental data. However, since the model~(\ref{KP}) is phenomenological in nature, the accuracy of the found resonant frequencies and ways for improvement are unclear. In addition, the model does not not account for the linewidth of the plasmon resonances, including the radiative contribution, nor does it provide information on the final line shape.

\begin{figure}[!t]
    \centering
    \includegraphics[width=\linewidth]{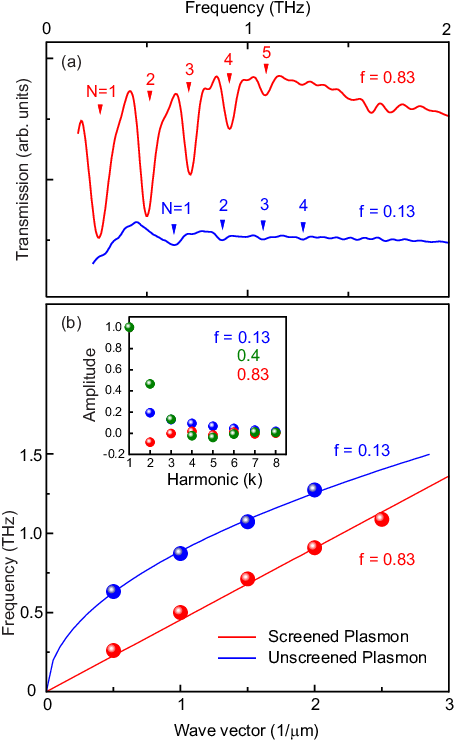}
    \caption{(a) The transmission spectra recorded for the plasmonic crystal samples with a period of $p=12$~\textmu{}m and different filling factors $f=0.83$ (red curve) and $f=0.13$ (blue curve). For each of the curves, the frequency positions of the successive plasma resonances are labeled as $N=1,2,3 \ldots$ (b) The dispersions of the plasma excitations obtained under the assumption of $q=N \times 2 \pi/p$. The red line shows a linear dispersion law according to Eq.~(\ref{scr_plasmon}), which is characteristic of the screened plasmon. The blue line presents a square-root dispersion law according to Eq.~(\ref{plasmon}), which is characteristic of the unscreened plasmon. The inset shows the normalized intensity of the $k$-th space harmonic of the fundamental plasmon mode calculated using our theory for $f=0.13, 0.4$, and $0.83$.}
    \label{4}
\end{figure} 

In order to find a better way to describe the behavior of plasmon modes in the plasmonic crystal, we develop a method based on the solution of Maxwell's equations for space harmonics of the electric and magnetic fields:
\begin{equation}
\begin{split}
& E_{x,z}(x,z,t)=\sum_{k=-\infty}^{\infty} E_{x,z}^{k}(z) e^{iq_k x-i\omega t},\\ 
& H_y(x,z,t)=\sum_{k=-\infty}^{\infty} H_y^{k}(z) e^{iq_k x-i\omega t},
\end{split}
\end{equation}
where $q_k=2\pi k/p$ is the $k$-th space harmonic of the grating, $z$-axis is oriented perpendicular to the 2DES, $y$-axis is oriented along the grating, and $x$-axis is perpendicular to the grating. The detailed derivation of the dispersion equation below can be found in the Supplemental Material~\cite{Supplemental}. When describing the plasmons, we consider the gates to be ideal metals and assume that currents are negligible at the edges. Additionally, we note that the wavelength of the THz radiation is significantly larger than the period of the plasmonic crystal, which substantially simplifies our analysis. Lastly, given that the plasmon decay rate, caused by collisions with impurities, phonons, and the emission of electromagnetic waves, is minimal, we can derive the following equation for the fundamental plasmon frequency, $\omega_p$:
\begin{equation}
\label{eq:theor_freq}
    g(f)+\sum_{k=1}^{+\infty} \frac{ k^2(1+\varkappa)e^{-4\pi h k/p}}{\omega_p^2/\omega_u^2-k(1+\varkappa e^{-4\pi h k/p})}  \left(\frac{\cos \pi k f }{1-4 k^2f^2}\right)^2=0,
\end{equation}
where $\omega_u=\sqrt{2\pi \,  n_s e^2/2 m^{\ast}  \varepsilon_{\rm GaAs} \varepsilon_0 p }$ is the frequency of unscreened plasmon~(\ref{plasmon}) at the wave vector $q_u=2\pi/p$, $\varkappa=(\varepsilon_{\rm GaAs}-1)/(\varepsilon_{\rm GaAs}+1)$, and the function $g(f)$ is defined via the series
\begin{equation}
    g(f)=\sum_{k=1}^{+\infty} k \left(\frac{\cos \pi k f }{1-4 k^2f^2}\right)^2
\end{equation}
and can be expressed through the special functions. It is important to note that each plasmon mode of the plasmonic crystal consists of a superposition of standing plasma waves with wave-vectors $q_k=2\pi k/p$ propagating through the plasmonic crystal. To achieve good convergence, we left a sufficiently large number of space harmonics (one hundred) in the sum of Eq.~(\ref{eq:theor_freq}), after which the desired plasmon frequency can be calculated numerically. The resulting frequencies of the fundamental plasmon mode, calculated from Eq. (\ref{eq:theor_freq}), are shown as solid curves in Fig.~\ref{3} for different filling factors. There is a remarkable agreement between theoretical predictions and experimental results, which supports the validity of our theoretical approach.

Several important observations can be made. It is particularly interesting that the plasmon frequency remains essentially constant, starting from the fully screened case of $f=1.0$ and all the way down to $f=0.6$. In this state, the frequency of the plasmon resonance stays close to the frequency of the screened plasmon Eq.\,(\ref{scr_plasmon}) with $q=2 \pi/p$ (the empty arrows in Fig.~\ref{3}). Beyond this point, the plasmon frequency increases sharply and eventually saturates near the filling factor of $f=0.1$, aligning with the frequency of the unscreened plasmon with $q=2 \pi/p$, Eq.\,(\ref{plasmon}), and the dielectric permittivity  $\varepsilon(q) = 7.7$ for $p=12$~\textmu{}m and $\varepsilon(q) = 8.0$ for $p=8$~\textmu{}m  (arrows in Fig.~\ref{3}). Notably, the calculated values of the effective dielectric permittivity are slightly larger than the average value of $(1 + \varepsilon_{\rm GaAs})/2 = 6.9$. This difference arises due to the $0.22$\,\textmu{}m GaAs top-layer covering the 2DES~\cite{Supplemental}.

To further illustrate the superlattice nature of the plasmon modes excited in the plasmonic crystal, we plot the normalized intensity of the calculated $k$-th space harmonic of the fundamental plasmon mode for $f=0.13, 0.4$, and $0.83$ (inset of Fig.~\ref{4}). The normalization is done with respect to the intensity of the $k=1$ fundamental harmonic. For small ($f=0.13$) and large ($0.83$) filling factors, the fundamental superlattice harmonic $k=1$ dominates the plasmonic response. Whereas, for intermediate $f$-values, the  $k=2$ and $3$ harmonics become significant. Indeed, Fig.~\ref{4}(a) shows the spectrum of the plasma excitations observed for the sample with $l_g=9.9$~\textmu{}m and period $p=12$~\textmu{}m ($f=0.83$). We now take the frequencies of the successive plasma peaks and assign them sequential wave vectors $q=N \times 2 \pi/p$ ($N=1,2,3 \ldots$). Here, we assume that only $k=1$ space harmonic contributes to the plasmonic response. The theoretical prediction, shown as a red line in Fig.~\ref{4}(b), exhibits a linear form characteristic of the screened plasmon. The theory (\ref{scr_plasmon}) shown by a red line in Fig.~\ref{4}(b) adequately agrees with the experimental points. In the opposite limit of narrow gates, when $l_g=1.6$~\textmu{}m ($f=0.13$), the dispersion obtained using the same procedure has a square-root dependency on $q=N \times 2 \pi/p$ (blue dots in Fig.~\ref{4}(b)). This behavior is typical of the unscreened plasmon, with its dispersion described by Eq.\,(\ref{plasmon}), represented by a blue line in Fig.~\ref{4}(b).  In the most common case of intermediate gate widths, the fundamental plasmon state arises from the hybridization of different superlattice space harmonics. Thus, a simple expression for the plasmon frequency cannot be provided.

\begin{figure}[!t]
    \centering
    \includegraphics[width=\linewidth]{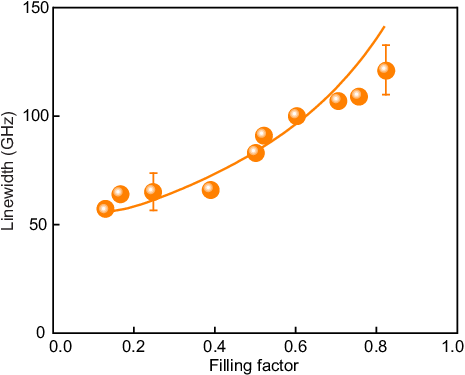}
    \caption{The linewidth of the fundamental plasma mode versus the filling factor $f = l_g/p$. The sample with the period of the plasmonic crystal $p = 12$~\textmu{}m. The orange line shows the result of our theoretical analysis~\cite{Supplemental}.}
    \label{5}
\end{figure} 

Another matter of interest is associated with the relaxation dynamics of plasmons in the plasmonic crystal. Figure~\ref{5} shows the dependence of the fundamental plasma mode linewidth, $\Delta \nu$, on the filling factor. For small filling factors $f$, the resonance linewidth is primarily influenced by a single-particle scattering. Indeed, a linewidth of $\Delta \nu = 54$~GHz corresponds to a relaxation time of $\tau = 1/(2\pi \Delta \nu) = 3$~ps, which agrees with the electron mobility obtained from the transport measurements. However, as the filling factor increases, we observe a significant broadening of the resonance, accompanied by an increase in amplitude. This suggests that the plasmon linewidth is influenced by an additional mechanism beyond just single-particle scattering. The solid line in Fig.~\ref{5} shows the result of the theoretical analysis. 2D plasmons on a homogeneous surface are non-radiative modes. Nevertheless, the gate grating used for the coupling process also couples electromagnetic radiation out of the plasmon excitation~\cite{Tsui:1980, Heitmann:1982}. Such radiative damping is typical for plasmon excitations when the plasmon wavelength is comparable to the wavelength of the incident electromagnetic wave~\cite{Andreev:2015, Muravev:2016}. However, the observation of radiative effects for the plasmon excitations in the $p \ll \lambda$ limit is quite surprising.

To gain some feeling about the strength of the radiative decay in our system, we used the simplest expression for the linewidth of the plasmon resonance in 2DES~\cite{Mikhailov:1996, Mikhailov:2004}:
\begin{equation}
\Delta \omega = \frac{1}{\tau} + \Gamma_{\rm rad} = \frac{1}{\tau} + \frac{n_s e^2}{m^{\ast} c \, \varepsilon_0 (1+ \sqrt{\varepsilon_{\rm GaAs}})}.
\end{equation}
This expression provides an estimate of the maximum radiative decay rate that we can expect for our plasmonic crystal, $\Delta \nu_{\rm rad} = \Gamma_{\rm rad}/2 \pi = 65$~GHz. As illustrated Fig.~\ref{5}, the experimental data demonstrate an increase in radiative damping with an increase in filling factor $f$. The range of values extends from $60$~GHz ($f=0.13$) to a maximum of $120$~GHz for a filling factor of $f=0.83$. Therefore, the gate grating performs as an efficient radiative antenna. 

In conclusion, we have investigated the plasma excitations in the terahertz transmission of the plasmonic crystal made from 2DES in the AlGaAs/GaAs semiconductor heterostructure with a periodic metal grating deposited on the surface. Experiments have demonstrated that the plasmonic crystal supports a series of plasmon modes, which evolve from an unscreened state in a narrow-gate configuration to a screened state for wide gates. An analytical approach has been developed that has managed to accurately describe the fundamental plasma state in the plasmonic crystal. It has been established that plasmons in a plasmonic crystal have a delocalized superlattice nature, being the combination of both screened and unscreened plasmons. Moreover, they consist of a superposition of multiple harmonics with wave vectors  $q_k=2\pi k/p$ ($k=1,2, 3 \ldots$). The obtained results establish a scientific foundation for engineering new plasmonic terahertz devices.

\begin{acknowledgments}
    The authors acknowledge the financial support from the Russian Science Foundation (Grant No.~19-72-30003). The authors are grateful to V.V.~Popov for discussing the results of the experiment at an early stage of the project.
\end{acknowledgments}

\bibliography{main}

\end{document}


\title
{Supplemental Material for 'On the Nature of the Fundamental Plasma Excitation in a Plasmonic Crystal'}

\author{A.~R.~Khisameeva$^{a}$, A.~Shuvaev$^{a}$, A.~A.~Zabolotnykh$^{b}$, A.~S.~Astrakhantseva$^{c}$, D.~A.~Khudaiberdiev$^{a}$, A.~Pimenov$^{a}$, I.~V.~Kukushkin$^{c}$, V.~M.~Muravev$^{c}$ \footnote{Corresponding E-mail: muravev@issp.ac.ru}}
\affiliation{$^a$ Institute of Solid State Physics, Vienna University of Technology, 1040 Vienna, Austria \\
$^b$ Kotelnikov Institute of Radio-engineering and Electronics of the RAS, Mokhovaya 11-7, Moscow 125009, Russia\\
$^c$ Institute of Solid State Physics, RAS, Chernogolovka, 142432 Russia}

\date{\today}\maketitle

\section{\textrm{I}. Terahertz Time-domain spectroscopy technique}

 The experiments were conducted using the time-domain spectroscopy setup, which covers a frequency range of $0.15$ to $2$~THz, see Fig. 1(c) of the main text. The system consists of two main parts. THz part, where the beam—generated and detected by photoconductive antennas acting as Hertzian dipoles—interacts with the sample. The optical part utilizes a femtosecond laser to trigger and measure the THz pulse. By adjusting the delay of the probe laser, a complete time-resolved signal can be acquired. The spectrometer is equipped with a continuous-flow liquid helium cryostat with transparent windows. All measurements are conducted at the base sample temperature of $T=5$~K. To minimize noise from water absorption lines in the air, the THz beam path is enclosed in a chamber that can be purged with nitrogen or evacuated. Spectroscopy measurements are carried out using a  $6$~mm aperture positioned close to the sample, ensuring that the electromagnetic radiation transmits only through the grating-gate active region of the plasmonic crystal. The polarization of the electromagnetic radiation is directed perpendicular to the grids.

\section{\textrm{II}. Detailed description of all measured samples}

The information on all the samples under study is presented in the tables. Each table corresponds to a specific grating period of either $8$ or $12$ $\mu$m. The samples with an active area of $8 \times 8$~mm were fabricated using standard photolithography, while those measuring $5 \times 5$~mm were made using electron beam lithography.

\begin{table}[h]
    \centering
    \begin{tabular}{|l|c|c|c|c|c|c|c|c|c|c|c|c|c|c|c|}
        \hline
        \textbf{Structure parameters} & \textbf{S1} & \textbf{S2} & \textbf{S3} & \textbf{S4} & \textbf{S5} & \textbf{S6} & \textbf{S7} & \textbf{S8} & \textbf{S9} & \textbf{S10} & \textbf{S11} & \textbf{S12} & \textbf{S13} & \textbf{S14} & \textbf{S15} \\
        \hline
        Grid period $p$, $\mu$m & 11.8 & 11.8 & 11.9 & 11.9  & 11.7 & 12.0  & 12.0  & 12.0 & 12.1 & 11.9  & 11.9  & 12.0  & 12.1  & 12.0  & 12.0 \\
        \hline
        Gated region width $l_g$, $\mu$m & 11.4  & 11.3 & 11.3 & 10.7 & 10.2 & 9.9  & 9.1 & 8.5 & 7.2 & 6.3 & 6.1 & 4.7 & 3.0 & 2.0 & 1.6 \\
        \hline
        Ungated region width $l_u$, $\mu$m & 0.4 & 0.5 & 0.6 & 1.2 & 1.5  & 2.1 & 2.9 & 3.5 & 4.9 & 5.7 & 5.8  & 7.3  & 9.1  & 10.0  & 10.4 \\
        \hline
        Filling factor $f$ & 0.97  & 0.96 & 0.95 & 0.90  & 0.87 & 0.83  & 0.76 & 0.71 & 0.60 & 0.53  & 0.51  & 0.39  & 0.25  & 0.17  & 0.13 \\
        \hline
        Active area, mm$^2$ & \multicolumn{5}{c|}{6 $\times$ 6} & \multicolumn{9}{c|}{8 $\times$ 8} & 5 $\times$ 5\\
        \hline
    \end{tabular}
    \caption{Plasmonic crystal parameters for samples with grating period of $p=12$ $\mu$m}
    \label{tab:structure_parameters_p12}
\end{table}

\begin{table}[h]
    \centering
    \begin{tabular}{|l|c|c|c|c|c|c|c|c|c|c|c|c|c|c|c|c|c|}
        \hline
        \textbf{Structure parameters} & \textbf{S1} & \textbf{S2} & \textbf{S3} & \textbf{S4} & \textbf{S5} & \textbf{S6} & \textbf{S7} & \textbf{S8} & \textbf{S9} & \textbf{S10} & \textbf{S11} & \textbf{S12} & \textbf{S13} & \textbf{S14} & \textbf{S15} & \textbf{S16} & \textbf{S17}\\
        \hline
        Grid period $p$, $\mu$m & 8.0  & 8.0  & 8.0 & 8.0 & 8.0  & 8.0  & 8.0  & 8.0  & 8.0  & 8.0 & 8.0  & 8.0  & 8.0 & 8.0 & 8.0  & 8.0  & 8.0  \\
        \hline
        Gated region width $l_g$, $\mu$m & 6.6  & 5.8  & 5.0 & 4.2 & 3.8  & 3.5  & 3.4  & 3.3  & 2.8  & 2.4 & 2.3 & 2.1 & 1.5  & 1.7 & 1.2 & 1.0  & 0.8   \\
        \hline
        Ungated region width $l_u$, $\mu$m & 1.4  & 2.2  & 3.0 & 3.8 & 4.2  & 4.5  & 4.6  & 4.7  & 5.2  & 5.6 & 5.7  & 5.9  & 6.5 & 6.3 & 6.8  & 7.0  & 7.2 \\
        \hline
        Filling factor $f$ & 0.83  & 0.73  & 0.63 & 0.53 & 0.48  & 0.44  & 0.43  & 0.41  & 0.35  & 0.30 & 0.29  & 0.26  & 0.19 & 0.21 & 0.15  & 0.13  & 0.1\\
        \hline
        Active area, mm$^2$ & \multicolumn{13}{c|}{8 $\times$ 8} & \multicolumn{4}{c|}{5 $\times$ 5} \\
        \hline
    \end{tabular}
    \caption{Plasmonic crystal parameters for samples with grating period of $p=8$ $\mu$m}
    \label{tab:structure_parameters_p8}
\end{table}

\section{\textrm{III}. Additional transmission spectra for the plasmonic crystal}

\begin{figure}[h!]
    \centering
    \includegraphics[width=0.65\linewidth]{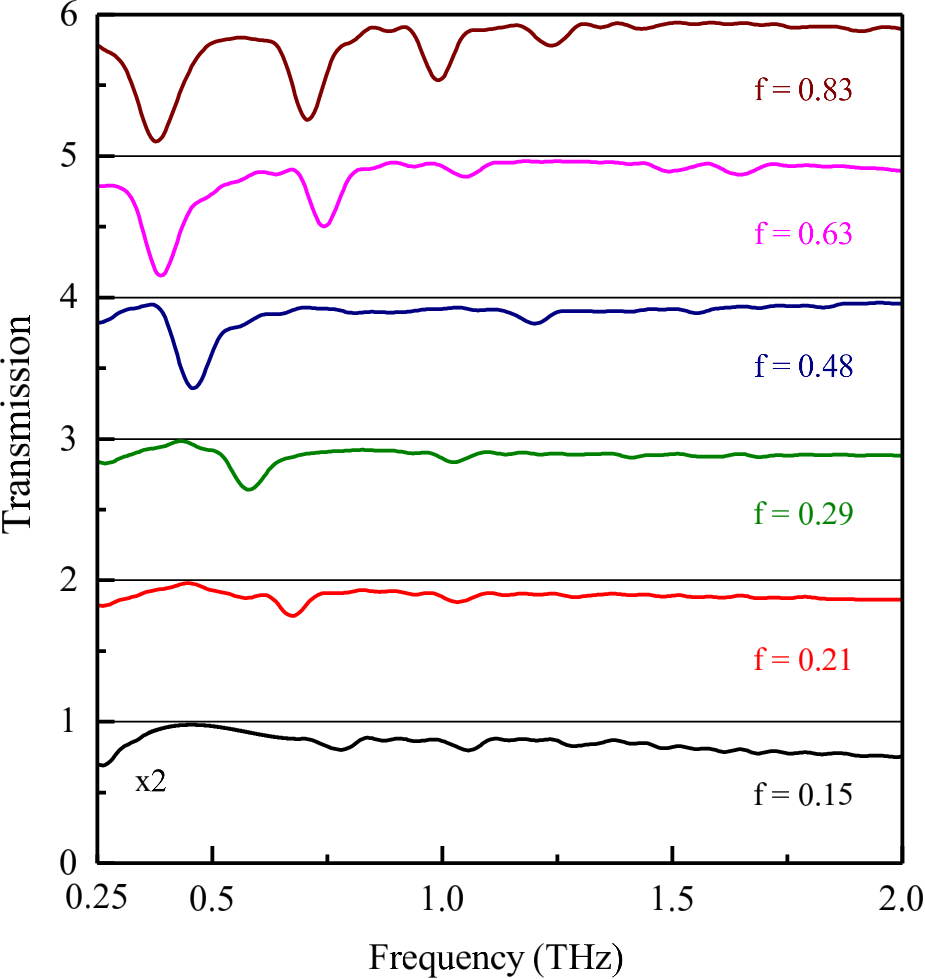}
    \caption{Transmission spectra measured for the plasmonic crystal with different grid filling factors. The grid period is kept fixed at $p = 8$~\textmu{}m.}
    \label{S1}
\end{figure}

Figure~\ref{S1} shows the transmission through a series of samples with varying gate widths $l_g= 1.2, 1.7, 2.3, 3.8, 5.0$ and $6.6$~\textmu{}m, keeping the grid period constant at $p = 8$~\textmu{}m. For clarity, the curves are shifted along the vertical axis. The polarization of the incident electromagnetic wave is directed perpendicular to the grid direction.

\newpage

\section{\textrm{IV}. Analysis of the linewidth}

\begin{figure}[h!]
    \centering
    \includegraphics[width=0.9\linewidth]{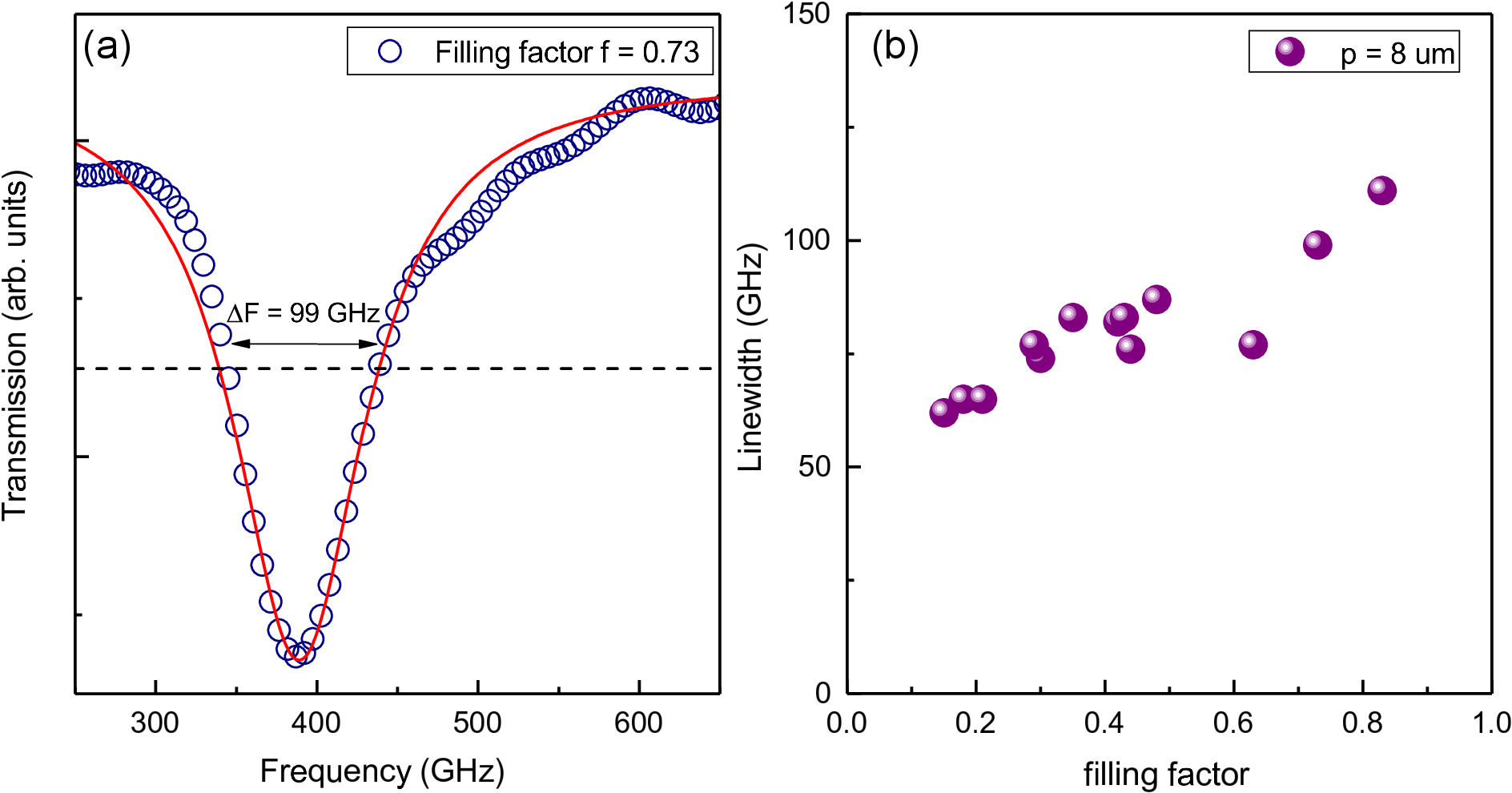}
    \caption{(a) Width analysis for sample with filling factor $f=0.73$: blue circles represent the experimental data, red line - Lorentzian fit. (b) The linewidth of the fundamental plasma mode versus the filling factor $f = l_g/p$. The grid period is kept fixed at $p = 8$~\textmu{}m.}
    \label{S_Fit}
\end{figure}

Figure~\ref{S_Fit}(a) shows the linewidth extracted from the Lorentz fit of the fundamental plasmon resonance for the sample with the period of the plasmonic crystal $p = 8$~$\mu$m and filling factor of $f = 0.73$. The blue circles represent the experimental transmission spectrum, the red line shows the Lorentzian fit, and the dashed line marks "half-height" of the resonance.

The resultant dependence of the fundamental plasma mode linewidth  on the filling factor $f$ are demonstrated in Figure~\ref{S_Fit}(b), the resultant dependence of the fundamental plasma mode linewidth,

\newpage

\section{\textrm{V}. The Kronig-Penney model for plasmon modes in the plasmonic crystal}

In the Kronig-Penney model, the dispersion relation for a grating plasmonic crystal is derived using the following assumptions~\cite{AizinS, PopovS, KachorovskiiS, SvintsovS, KachorovskiiS}:

\begin{enumerate}
\item Plane-wave expansion of the electric potential $\varphi (x) \sim e^{i qx - i \omega t}$. Here dispersion $\omega (q)$ is determined by the local dispersions in each region of the 2DES.

\item Matching of the potential and electric current at the gated - ungated boundaries.

\item  Imposing Bloch periodicity conditions for the fields in the adjacent gratings, $\varphi(x +
p) = \varphi(x) e^{i \Theta}$ ($\Theta =0$ for normal incidence).
\end{enumerate}

In general, the geometry of the sample has two symmetry planes, one in the middle of the gated region and one in the middle of the ungated region, see Fig.~\ref{S2}. In this case, solutions can be sought in the form of symmetric and antisymmetric functions. We first consider the mode with antisymmetric potential $\varphi$ (symmetric current density $j$):
\begin{equation}
\begin{aligned}
    &\varphi_g = A_1 \sin(q_g x), \qquad -l_g/2 < x < l_g / 2 \\
    &\varphi_u = A_2 \sin\left(q_u \left(x - \frac{l_g + l_u}{2} \right) \right), \qquad p/2 > x > l_g / 2 \\
    &\left(\varphi_u = A_2 \sin\left(q_u \left(x + \frac{l_g + l_u}{2} \right) \right), \qquad -p/2 < x < -l_g / 2 \right) \\
\end{aligned}
\label{phi_asym}
\end{equation}
The last expression for $\varphi_u$ in the region $x < -l_g / 2$ follows from the periodic boundary condition $\varphi(x + p) = \varphi(x)$ with $\Theta=0$. Now, we relate the current density $j$ with the potential $\varphi$. From the definition of the electric potential $\vect{E} = -\vect{\nabla} \varphi$, and the Ohm's law $\vect{j} = \sigma \vect{E}$, we have:
\begin{equation}
j = \sigma E = -\sigma \frac{d \varphi}{dx}.
\end{equation}
The current density for the $j$-symmetric mode is then:
\begin{equation}
\begin{aligned}
    &j_g = -\sigma A_1 q_g \cos(q_g x), \qquad -l_g/2 < x < l_g / 2 \\
    &j_u = -\sigma A_2 q_u \cos\left(q_u \left(x - \frac{l_g + l_u}{2} \right) \right), \qquad p/2 > x > l_g / 2 \\
&\left(j_u = -\sigma A_2 q_u \cos\left(q_u \left(x + \frac{l_g + l_u}{2} \right) \right), \qquad -p/2 < x < -l_g / 2 \right) \\
\end{aligned}
\label{j_sym}
\end{equation}
We match both potential $\varphi$ and current density $j$ at $x = l_g / 2$ (matching at $x = -l_g / 2$ is fulfilled automatically with the choice~(\ref{phi_asym}) and (\ref{j_sym})):
\begin{equation}
\begin{aligned}
A_1 \sin\left(\frac{q_g l_g}{2} \right) &= -A_2 \sin\left(\frac{q_u l_u}{2} \right) \\
-\sigma A_1 q_g \cos\left(\frac{q_g l_g}{2} \right) &= -\sigma A_2 q_u \cos\left(\frac{q_u l_u}{2} \right).
\end{aligned}
\end{equation}
In order to have nontrivial solutions, the determinant of this system should be zero, which gives the dispersion relation for the $j$-symmetric mode:
\begin{equation}
q_u \sin\left(\frac{q_g l_g}{2} \right) \cos\left(\frac{q_u l_u}{2} \right) + q_g \cos\left(\frac{q_g l_g}{2} \right) \sin\left(\frac{q_u l_u}{2} \right) = 0.
\label{active_mode}
\end{equation}
We now determine the charge density $\rho$ which is connected to the current density $j$ via the continuity equation:
\begin{equation}
\frac{\partial \rho}{\partial t} + \dive \vect{j} = 0; \qquad \frac{\partial \rho}{\partial t} + \frac{dj}{dx} = 0; \qquad -i \omega \rho + \frac{dj}{dx} = 0; \qquad i \omega \rho = \frac{dj}{dx}.
\end{equation}
We have used the implied time dependency of all quantities $\sim e^{-i \omega t}$. The charge densities in gated and ungated regions are then:
\begin{equation}
\begin{aligned}
    &i \omega \rho_g = \sigma A_1 q_g^2 \sin(q_g x), \qquad -l_g/2 < x < l_g / 2 \\
    &i \omega \rho_u = \sigma A_2 q_u^2 \sin\left(q_u \left(x - \frac{l_g + l_u}{2} \right) \right), \qquad x > l_g / 2 \\
    &\left(i \omega \rho_u = \sigma A_2 q_u^2 \sin\left(q_u \left(x + \frac{l_g + l_u}{2} \right) \right), \qquad x < -l_g / 2 \right) \\
\end{aligned}
\label{rho_asym}
\end{equation}
The distribution of charge density $\rho$ is antisymmetric relative to the middle of the gated or ungated regions, implying the existence of nonzero dipole moment in both the gated and ungated zones. This means that this mode couples to the electric field of the incident radiation and can be optically excited.
The symmetric distribution of the current density $j$ is also compatible with the excitation by the homogeneous electric field of the incident radiation.

\begin{figure}[htbp]
    \centering
    \includegraphics[width=0.65\linewidth,clip]{kp_modes.eps}
    \caption{The frequency of the fundamental plasma mode in the plasmonic crystal as a function of the filling factor $f = l_g / p$ according to the Kronig-Penney model~(\ref{active_mode}) and (\ref{dark_mode}) with simplified relations~(\ref{Supp:Disp_simplified}). The grid period is kept fixed at $p = 12$~\textmu{}m. Solid black line is the optically active mode, red dashed line denotes the dark mode.}
    \label{kp_modes}
\end{figure}

In contrast, the mode with antisymmetric current $j$ (symmetric charge density $\rho$) cannot be excited optically and can be designated as a dark mode. The derivation of the dispersion relation of the dark mode is similar. Its potential is:
\begin{equation}
\begin{aligned}
    \varphi_g &= B_1 \cos(q_g x), \qquad -l_g/2 < x < l_g / 2 \\
    \varphi_u &= B_2 \cos\left(q_u \left(x - \frac{l_g + l_u}{2} \right) \right), \qquad p/2 > x > l_g / 2 \\
\end{aligned}
\label{phi_sym}
\end{equation}
the current density:
\begin{equation}
\begin{aligned}
j_g &= \sigma B_1 q_g \sin(q_g x), \qquad -l_g/2 < x < l_g / 2 \\
j_u &= \sigma B_2 q_u \sin\left(q_u \left(x - \frac{l_g + l_u}{2} \right) \right), \qquad p/2 > x > l_g / 2 \\
\end{aligned}
\label{j_asym}
\end{equation}
and the charge density:
\begin{equation}
\begin{aligned}
    i \omega \rho_g &= \sigma B_1 q_g^2 \cos(q_g x), \qquad -l_g/2 < x < l_g / 2 \\
    i \omega \rho_u &= \sigma B_2 q_u^2 \cos\left(q_u \left(x - \frac{l_g + l_u}{2} \right) \right), \qquad p/2 > x > l_g / 2 \\
\end{aligned}
\label{rho_sym}
\end{equation}
The dispersion relation of the dark mode is obtained by matching potential $\varphi$ and current density $j$ at $x = l_g / 2$:
\begin{equation}
q_u \sin\left(\frac{q_u l_u}{2} \right) \cos\left(\frac{q_g l_g}{2} \right) + q_g \sin\left(\frac{q_g l_g}{2} \right) \cos\left(\frac{q_u l_u}{2} \right) = 0.
\label{dark_mode}
\end{equation}

\begin{figure}[htb]
    \centering
    \includegraphics[width=0.32\linewidth,clip]{kp_fields_f0p13.eps}
    \includegraphics[width=0.32\linewidth,clip]{kp_fields_f0p4.eps}
    \includegraphics[width=0.32\linewidth,clip]{kp_fields_f0p83.eps}
    \caption{Upper panels show the current density in 2DES, $j$, as a function of coordinate $x$ for both optically active (black solid curves) and dark (red dashed curves) modes. Different filling factors $f=l_g/p$ are: $0.13$ (left panels), $0.4$ (middle panels), and $0.83$ (right panels). Lower panels show the charge density in 2DES, $\rho$, for the same filling factors. The gated part of 2DES extends from $-f/2$ to $f/2$ in dimensionless coordinates $x/p$, the boundaries are shown by vertical gray lines.}
    \label{kp_fields}
\end{figure}

In order to solve Eqs.~(\ref{active_mode}) and (\ref{dark_mode}) we recall that $l_u=p-l_g$, values of $q_u$ and $q_g$ are related with the frequency of the incident wave, $\omega$, via the dispersion equations for unscreened and screened plasmons. Namely, in the case of our system (in the limit $\omega \tau \gg 1$ and neglecting the electromagnetic retardation effects), the relations are as follows:
\begin{equation}
\label{Supp:KPadd}
    \omega = \sqrt{\frac{n_s e^2 q_u}{2 m^{\ast} \varepsilon_{\rm GaAs} \varepsilon_0} \left(1+ \frac{\varepsilon_{\rm GaAs}-1}{\varepsilon_{\rm GaAs}+1} e^{-2q_u h}\right)} \quad \text{and} \quad
    \omega = \sqrt{\frac{n_s e^2 q_g}{2 m^{\ast} \varepsilon_{\rm GaAs} \varepsilon_0} \left(1- e^{-2q_g h} \right)}.   
\end{equation}
The result obtained with the use of Eqs.~(\ref{active_mode}) and (\ref{Supp:KPadd}) is presented in Fig.~3 of the main text by the dashed lines. Figure~\ref{kp_modes} shows both the optically active (solid black curve) and dark (red dashed curve) modes calculated for the period of the plasmonic crystal $p = 12$~\textmu{}m. Here, simplified (at $q_{g,u}h \ll 1$) dispersion relations for gated and ungated plasmons were used:
\begin{equation}
\label{Supp:Disp_simplified}
    \omega = \sqrt{\frac{n_s e^2 q_u}{m^{\ast} (\varepsilon_{\rm GaAs} + 1)\varepsilon_0}} \quad \text{and} \quad
    \omega = \sqrt{\frac{n_s e^2 h}{m^{\ast} \varepsilon_{\rm GaAs} \varepsilon_0}} \ q_g.
\end{equation}
The parameters of the sample are $n_s = 1.47 \times 10^{12}$~cm$^{-2}$, $m^{\ast} = 0.076 \, m_0$, $\varepsilon_{\rm GaAs} = 12.8$, and $h = 0.22$~\textmu{}m.
The upper panels of Fig.~\ref{kp_fields} demonstrate the distribution of the current density $j$ along the period of the plasmonic crystal for three different filling factors $f = l_g / p$. The charge density $\rho$ is shown in the lower panels. Again, both optically active (solid black curves) and dark (red dashed curves) modes are shown.

Note that the total charge under the gated region is nonzero for the dark modes, which can be best seen for the filling factor $f = 0.13$ shown in the lower left panel of Fig.~\ref{kp_fields}. This means that an external electrode between 2DES and the gate is required in order to excite the dark mode. Pure optical excitation of the sample with isolated stripes is not possible.

\section{\textrm{VI}. Analytical description of plasmons in 2DES with metal grating}

\begin{figure}[htb]
    \centering
    \includegraphics[width=0.5\linewidth]{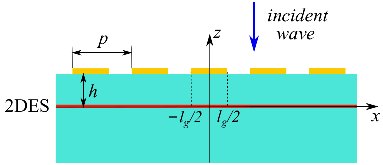}
    \caption{Schematic of the structure under study.}
    \label{S2}
\end{figure}

Consider 2DES placed at $z=0$, see Fig.~\ref{S2}, while metal grating is situated at $z=h$ and grid strips are assumed to be $\delta$-thin in $z$-direction. The dielectric permittivity $\varepsilon(z)$ equals unity at $z>h$ and $\varepsilon$ at $z<h$, which corresponds to GaAs substrate of infinite thickness. The system is periodic in the $x$-direction and uniform along the $y$-axis. The external electromagnetic wave falls perpendicular to the system with the electric field polarized in the $x$-direction.

In general, such a layered system can support two types of electromagnetic excitations: so-called TM and TE modes. The former have components of electric and magnetic fields $(E_x,H_y,E_z)$, while the latter have components $(H_x,E_y,H_z)$. As in the experimental setup the incident electric field is polarized in the $x$-direction, only TM modes are excited and, consequently, we focus below on the analysis of the response due to these particular modes.

To find the electromagnetic response of the system, we use an approach similar to the one introduced in Ref.~\cite{MikhailovS}. We start with the solution of the Maxwell's equations, which (in CGS units) appear as follows:
\begin{equation}
\label{Supp:Maxeq}
    rot { \bf H}=\frac{\varepsilon(z)}{c} \frac{\partial {\bf E} }{\partial t } +\frac{4 \pi }{c} {\bf j}, \quad rot {\bf E}=-\frac{1}{c} \frac{\partial {\bf H} }{\partial t },
\end{equation}
where $c$ is the speed of light in vacuum, ${\bf j}={\bf j}(x,z,t)$ is the charge current density in 2DES ($z=0$) and metal gates ($z=h$).

We look for periodic in $x$-directions solutions, so we introduce the following Fourier series for the sought functions:
\begin{equation}
    E_{x,z}(x,z,t)=\sum_{k=-\infty}^{\infty} E_{x,z}^{k}(z) e^{iq_k x-i\omega t}, \,\, H_y(x,z,t)=\sum_{k=-\infty}^{\infty} H_y^{k}(z) e^{iq_k x-i\omega t}, \,\, j_x(x,z,t)=\sum_{k=-\infty}^{\infty} j_x^{k}(z) e^{iq_k x-i\omega t},
\end{equation}
where $q_k=2\pi k/p$ is the $k$-th 'wave vector' of the grating, $p$ is the period of the structure, and $\omega$ is the frequency of the wave.

Further, we exclude from Eqs.~(\ref{Supp:Maxeq}) fields $H_y^k(z)$ and $E_z^k(z)$:
\begin{equation}
    H_y^k= \frac{c}{i \omega} \left( \partial_z E_x^k -i q_k E_z^k\right), \quad
    E_z^k=\frac{-i q_k}{q_k^2-\omega^2 \varepsilon(z)/c^2} \partial_z E_x^k,
\end{equation}
and obtain the single equation for $E_x^k(z)$ and $j_x^k(z)$:
\begin{equation}
\label{Supp:waveeq}
    \partial_z \frac{\varepsilon(z)}{q_k^2-\omega^2 \varepsilon(z)/c^2} \partial_z E_x^k(z) - \varepsilon(z) E_x^k(z) = \frac{4 \pi}{ -i \omega} j_x^k(z),
\end{equation}
where the density current appears as follows $j_x^k(z)=j_x^{k,2DES}\delta(z)+j_x^{k,metal}\delta(z-h)$. The resulting equation~(\ref{Supp:waveeq}) describes electromagnetic waves taking into account the inhomogeneity of the dielectric permittivity $\varepsilon(z)$ along the $z$-axis and the presence of currents in 2DES and metal grating.

We look for the solution of Eq.~(\ref{Supp:waveeq}) in the form
\begin{equation}
\label{Supp:form}
E_x^k(z)=
\begin{cases}
     C_1^k e^{-\beta^0_k z} +E_0 e^{-i\omega z/c} \delta_{k0}, & z\geq h; \\
     C_2^k e^{\beta_k z}+ C_3^k e^{-\beta_k z}, & h>z>0; \\
     C_4^k e^{\beta_k z}, & z\le 0;
\end{cases}
\end{equation}
where $\delta_{k0}$ is the Kronecker delta, which equals one at $k=0$ and zero at $k \neq 0$, term $E_0 e^{-i\omega z/c} \delta_{k0}$ describes the wave incident on the system against the $z$-axis (we assume $\omega>0$), $\beta_k=\sqrt{q_k^2-\omega^2 \varepsilon/c^2}$ and $\beta_k^0=\sqrt{q_k^2-\omega^2/c^2}$ determine how field $E_x^k(z)$ extends into dielectric medium and vacuum. Note that at $q_k<\omega \sqrt{\varepsilon}/c$ or $q_k<\omega/c$ coefficients $\beta_k$ or $\beta_k^0$, respectively, become imaginary with negative imaginary part, which corresponds to waves outgoing from the system.

To find the solution, along with Eq.~(\ref{Supp:form}), we need to impose boundary conditions for the electric field $E_x^k(z)$ at $z=0$ and $z=h$. The first condition is the continuity of $E_x^k(z)$, since it is tangent to the surfaces $z=0$ and $z=h$. The second boundary condition can be derived via the integration of Eq.~(\ref{Supp:waveeq}) near $z=0$ and $z=h$. So, the conditions are as follows:
\begin{equation}
\label{Supp:BC}
    \frac{\varepsilon(z)}{q_k^2-\omega^2 \varepsilon(z)/c^2} \partial_z E_x^k(z) |^{z=+0}_{z=-0}= \frac{4 \pi}{ -i \omega} j_x^{k, 2DES}, \quad  \frac{\varepsilon(z)}{q_k^2-\omega^2 \varepsilon(z)/c^2} \partial_z E_x^k(z) |^{z=h+0}_{z=h-0}= \frac{4 \pi}{ -i \omega} j_x^{k, metal}.
\end{equation}
In addition, since 2DES is homogeneous, one can use Ohm's law $j_x^{k,2DES}=\sigma E_x^k(z=0)$, where $\sigma=\sigma(\omega)$ is the dynamical conductivity of 2DES. Taking into account that $E_x^k(z=0)=C^k_4$, one can find from the boundary conditions at $z=0$ and continuity of $E_x^k(z)$ at $z=h$ the following relations:
\begin{equation}
\label{Supp:Coeffs}
    C^k_2=\left(1+\frac{2 \pi \sigma \beta_k}{-i\omega \varepsilon}\right)C^k_4, \quad
    C^k_3=\frac{2 \pi \sigma \beta_k}{i\omega \varepsilon} C^k_4, \quad 
    C^k_4=\frac{C_1^k e^{-\beta^0_k h} +E_0 e^{-i\omega h/c} \delta_{k0}}{\left(1+\frac{2 \pi \sigma \beta_k}{-i\omega \varepsilon}\right) e^{\beta_k h}+ \frac{2 \pi \sigma \beta_k}{i\omega \varepsilon} e^{-\beta_k h}}.
\end{equation}

Finally, using the relation $E_x^k(z=h)= C_1^k e^{-\beta^0_k h} +E_0 e^{-i\omega h/c} \delta_{k0}$ and the boundary condition (\ref{Supp:BC}) at $z=h$ the following equation for the electric field and the current density in a gate can be achieved:
\begin{equation}
\label{Supp:eq4E}
    E_x(x,z=h)=E_0 e^{-i\omega h/c}\frac{2}{\sqrt{\varepsilon}} T_{k=0}-\sum_{k=-\infty}^{+\infty} T_k \frac{4 \pi \beta_k}{-i \omega\varepsilon} \int_{-l_g/2}^{l_g/2} \frac{dx'}{p} j_x^{metal}(x') e^{iq_k (x-x')},
\end{equation}
where we have already taken into account that density current $j_x^{metal}(x)$ is zero outside the gate, $l_g$ is the width of a gate, and coefficients $T_k$ are defined as follows:
\begin{equation}
\label{Supp:eqTk}
    T_k=\frac{1+\frac{2\pi \sigma \beta_k}{-i\omega \varepsilon} \left(1-e^{-2\beta_k h}\right)}{\left(1+\frac{\beta_k}{\beta^0_k \varepsilon}\right)\left(1+\frac{2\pi \sigma \beta_k}{-i\omega \varepsilon}\right) + \left(1-\frac{\beta_k}{\beta^0_k \varepsilon}\right) \frac{2\pi \sigma \beta_k}{-i\omega \varepsilon} e^{-2\beta_k h}}.
\end{equation}
Note that the vanishing of the numerator of $T_k$ defines the dispersion equation for screened plasmon-polaritons~\cite{AndreevS}, while zeros of the denominator correspond to frequencies of electromagnetic waves in the system in the absence of the metal grid.

Eqs. (\ref{Supp:eq4E}) and (\ref{Supp:eqTk}) along with Ohm's law $j_x^{metal}(x)=\sigma_m E_x(x,z=h)$, where $\sigma_m$ is the conductivity of the metal gate, define the response of the system. It is important to note that the obtained equations follow directly from Maxwell's equations and the local Ohm's law without any approximations.

Now, we make several assumptions. First of all, we assume that the gates are ideal metals, so the electric field $E_x(x,z=h)$ in the left-hand side of Eq.~(\ref{Supp:eq4E}) is zero at $-l_g/2<x<l_g/2$, i.e. in the gates. The right-hand side of Eq.~(\ref{Supp:eq4E}) depends on the density current, $j_x^{metal}(x)$, which is nonzero only inside the gates. Therefore, Eq.~(\ref{Supp:eq4E}) can be solved in the interval $(-l_g/2,l_g/2)$ (where the left-hand side vanishes), with the desired function being $j_x^{metal}(x)$.

Secondly, we use the condition that the grid period is much less than the wavelength of the electromagnetic wave, i.e. $2\pi/p \gg \omega/c, \, \omega \sqrt{\varepsilon}/c$. Indeed, in the experimental setup $p=8$ and $12$~\textmu{}m, while the wavelength of radiation at $\nu=300$~GHz in GaAs is $c/ \nu \sqrt{\varepsilon_{GaAs}}\approx 280$~\textmu{}m. As a result, we take $\beta_k=\beta^0_k=|q_k|$ for $k \neq 0$, while $\beta_{k=0}=-i\omega\sqrt{\varepsilon}/c$ and $\beta_{k=0}^0=-i\omega/c$. In addition, we take that the distance $h$ is negligibly small compared to the wavelength of radiation.

Finally, to find the frequencies explicitly, we use the model function for the density current in the gate, namely, $j_x^{metal}(x)=j_A^m \cos(\pi x/l_g)$, which equals zero at the gate edges, $x=\pm l_g/2$. It should be noted that the used model function is more appropriate, especially for 'large' filling factors $l_g/p$ that are not much less than one, than the elliptic model applied in Ref.~\cite{MikhailovS}. This is likely due to the fact that trigonometric functions are eigen functions for the case of quasi-electrostatic plasmons in completely gated 2DES when conditions $l_g/p \to 1$ and $h \ll p$ are satisfied.

After using the model function for $j_x^{metal}(x)$, we multiply Eq.~(\ref{Supp:eq4E}) by $\cos(\pi x/ l_g)$ and integrate over the width of the strip $(-l_g/2,l_g/2)$. Then, the response of the system (as a function of frequency) can be determined through the amplitude of the current in the metal gates, $j_A^m$:
\begin{equation}
\label{Supp:eqresp}
    \frac{j_A^m}{E_0 e^{-i\omega h/c}} \frac{2\pi}{c} = \frac{\pi}{2f} \frac{-i\omega p(\varepsilon+1)}{2\pi c (\sqrt{\varepsilon}+1+ 4 \pi \sigma/c)} \frac{1}{\frac{-i\omega p(\varepsilon+1)}{2\pi c (\sqrt{\varepsilon}+1+ 4 \pi \sigma/c)}+ 2 \sum_{k=1}^{+\infty} k \frac{ \epsilon_{scr}(\omega,q_k)}{\epsilon_{uns}(\omega,q_k)}  \left(\frac{\cos \pi k f }{1-4 k^2f^2}\right)^2},
\end{equation}
where $\epsilon_{scr}(\omega,q_k)$ and $\epsilon_{uns}(\omega,q_k)$ are so-called effective dynamical permittivities of screened and unscreened 2DES: 
\begin{equation}
    \epsilon_{scr}(\omega,q)=1+\frac{2 \pi \sigma q}{-i \omega \varepsilon}\left(1-e^{-2qh}\right), \quad \epsilon_{uns}(\omega,q)=1+\frac{2 \pi \sigma q}{-i \omega \varepsilon}\left(1+ \frac{\varepsilon-1}{\varepsilon+1}e^{-2qh}\right).
\end{equation}

Let us analyze the denominator in the right-hand side of Eq.~(\ref{Supp:eqresp}). The first term of the denominator stems from the term with $k=0$ in the sum in Eq.~(\ref{Supp:eq4E}); it corresponds qualitatively to the relation of the incident electromagnetic wave with modes of the system and, consequently, to the radiative broadening of a resonance. The sum in the denominator of  Eq.~(\ref{Supp:eqresp}) describes the excitation of plasma waves, which in the case of our system are quasi-electrostatic excitations.

Now, to find the resonance frequency, the expression for the dynamic conductivity of 2DES should be specified. In what follows, we use the Drude model, namely, $\sigma=e^2 n_s /m^{\ast} (\tau^{-1}-i\omega)$, where $n_s$, $m^{\ast}$, and $\tau$ are respectively the electron concentration, the effective mass, and the relaxation time.

Considering the collisionless limit, $\omega \tau \gg 1$, and neglecting the first term in the denominator of Eq.~(\ref{Supp:eqresp}) one can find the resonance frequency from the condition of the vanishing of the denominator, which appears as follows:
\begin{equation}
\label{Supp:eqdisp}
    g(f)+\sum_{k=1}^{+\infty} \frac{ k^2(1+\varkappa)e^{-4\pi h k/p}}{\omega^2/\omega_u^2-k(1+\varkappa e^{-4\pi h k/p})}  \left(\frac{\cos \pi k f }{1-4 k^2f^2}\right)^2=0, \quad \text{where} \quad g(f)=\sum_{k=1}^{+\infty} k \left(\frac{\cos \pi k f }{1-4 k^2f^2}\right)^2,
\end{equation}
$\omega_u=\sqrt{2\pi n_s e^2 \, 2\pi / m^{\ast}  \varepsilon p }$ is the frequency of unscreened plasmons in 2DES embedded in dielectric medium with constant permittivity $\varepsilon$ at the wave vector $q=2\pi/p$ and $\varkappa=(\varepsilon-1)/(\varepsilon+1)$.

Terms in the sum of Eq.~(\ref{Supp:eqdisp}) decrease exponentially with the growth of $k$ due to the factor $e^{-4\pi h k/p}$. That is why, to find the frequency, we leave in the sum a sufficiently large number of terms, $k_{max}$, so that the condition $4\pi h k_{max}/p \gg 1$ is fulfilled, and then calculate the resonant frequency numerically. The found frequency as a function of the filling factor is presented in Fig.~3 of the main text, $k_{max}$ was equal to 100. 

To determine the linewidth of the resonance, we keep the first term in the denominator of Eq.~(\ref{Supp:eqresp}) and find the value of $|j_A^m 2\pi/E_0 c|$ as a function of frequency (here we also take $k_{max}=100$). Then, the resonance width at the $1/\sqrt{2}$ of the maximum corresponds to the desired linewidth. When calculating the linewidth, we also include into consideration the finite electron relaxation time, $\tau$, which was taken such that $2\pi/\tau=54$~GHz to harmonize the theory with the experimental data at small filling factors. The result obtained is presented in Fig.~5 of the main text.

Let us also give an analytical estimation of the resonant frequency by keeping only term $k=1$ in Eq.~(\ref{Supp:eqdisp}). Then, one can find 
\begin{equation}
\label{Supp:freq}
    \frac{\omega^2}{\omega_u^2}=1+\varkappa e^{-4 \pi h/p}- \frac{(1+\varkappa)e^{-4 \pi h/p}}{g(f)} \left(\frac{\cos \pi f }{1-4 f^2}\right)^2.
\end{equation}
The analysis shows that the obtained expression accurately describes the resonant frequency in the limit $f \to 0$; at $f>0$ the accuracy of Eq.~(\ref{Supp:freq}) is about 25\%. Consider in more detail the limit of the small filling factor $f \to 0$. In this case, $g(f) \to \infty$, and the frequency of the fundamental resonance coincides with the frequency of the unscreened plasmon mode:
\begin{equation}
    \omega=\omega_u \sqrt{ 1+\frac{\varepsilon-1}{\varepsilon+1} e^{-4\pi h/p}}= \sqrt{\frac{2 \pi n_s e^2 2 \pi}{m^{\ast} p } \frac{1}{\varepsilon(q=2 \pi/p)}},\quad \text{where} \quad \varepsilon(q)= \frac{\varepsilon}{1+\dfrac{\varepsilon-1}{\varepsilon+1} e^{-2qh } }
\end{equation}
corresponds to the effective permittivity, which depends on the 2D wave vector $q$. In the case of the experimental setup $h=0.22$~\textmu{}m, $\varepsilon=\varepsilon_{\rm GaAs}=12.8$, so for $p=8$~\textmu{}m and $p=12$~\textmu{}m we obtain $\varepsilon(q=2 \pi/p)$ to be equal to $8.0$ and $7.6$, correspondingly. Note that these values differ from the average dielectric permittivity between vacuum and GaAs substrate, i.e. $(\varepsilon_{\rm GaAs}+1)/2=6.9$, which appears in the case of $h=0$. Therefore, it is important to take into consideration the finite value of $h$, though $h \ll p$.

In the end, it should be noted that, due to the use of the model function for the density current in a gate, $j_x^{metal}(x)\propto \cos(\pi x/l_g)$, generally speaking, the applied method describes only those resonances whose excitation does not cause the current to become zero inside the metal strip, in particular, the fundamental resonance. To describe all resonances, further decomposition of the current $j_x^{metal}(x)$ into a series, for example, by trigonometric functions, should be done.

\section{\textrm{VII}. Distribution of electric field $E_x$ and charge density in 2DES}

Let us find the spatial distribution of electric field $E_x(x,z=0)$ and charge density, $\rho^{2DES}(x)$ in 2DES. Using Eq.~(\ref{Supp:eq4E}) and the expression for coefficient $C_4$ from Eq.~(\ref{Supp:Coeffs}) one can find the following relation:
\begin{multline}
\label{Supp:EF}
    E_x(x,z=0)=\frac{2 E_0 e^{i\omega h(\sqrt{\varepsilon}-1)/c}}{(\sqrt{\varepsilon}+1)(1+\frac{2\pi \sigma }{c\sqrt{\varepsilon}})+(\sqrt{\varepsilon}-1)\frac{2\pi \sigma }{c\sqrt{\varepsilon}}e^{2i \omega \sqrt{\varepsilon} h /c}} \\
    -\sum_{k=-\infty}^{+\infty} \frac{4 \pi \beta_k}{-i \omega\varepsilon}  \frac{j_x^{k, metal} e^{iq_k x-\beta_k h}}{\left(1+\frac{\beta_k}{\beta^0_k \varepsilon}\right)\left(1+\frac{2\pi \sigma \beta_k}{-i\omega \varepsilon}\right) + \left(1-\frac{\beta_k}{\beta^0_k \varepsilon}\right) \frac{2\pi \sigma \beta_k}{-i\omega \varepsilon} e^{-2\beta_k h}}.
\end{multline}

In what follows we do not analyze the homogeneous part on the field and focus of the inhomogeneous contribution to the field in 2DES (\ref{Supp:EF}), $E_x^{inh}(x)$. Making the same assumptions as when obtaining Eq.~(\ref{Supp:eqdisp}), namely, $\beta_k=q_k$ at $k \geq 1$, using the model function $j_x^{metal}(x)=j_A \cos(\pi x/l_g)$, and the collisionless Drude conductivity, one can find the following:
\begin{equation}
\label{Supp:ExInh}
    E_x^{inh}(x)=j_A 4f \frac{8 \pi }{-i \omega(\varepsilon+1) p} \sum_{k=1}^{+\infty} F(k) \cos 2 \pi k x , \quad \text{where} \quad F(k)= \frac{k e^{-2 \pi k h/p}}{\omega_u^2 k(1+\frac{\varepsilon-1}{\varepsilon+1} e^{-4\pi h k/p})/\omega^2-1} \cdot \frac{\cos \pi k f }{1-4 k^2f^2}
\end{equation}
are Fourier harmonics of the electric field $E_x^{inh}(x)$ and $\omega_u=\sqrt{2\pi n_s e^2 \, 2\pi / m^{\ast}  \varepsilon p }$ with $\varepsilon=\varepsilon_{\rm GaAs}$ in the case of the experimental setup under consideration. Values $F(k)/F(k=1)$ are presented in Fig.~4 of the main text for filling factors $f=0.13$, $0.4$, and $0.83$.

\begin{figure}[htb]
    \centering
    \includegraphics[width=\linewidth]{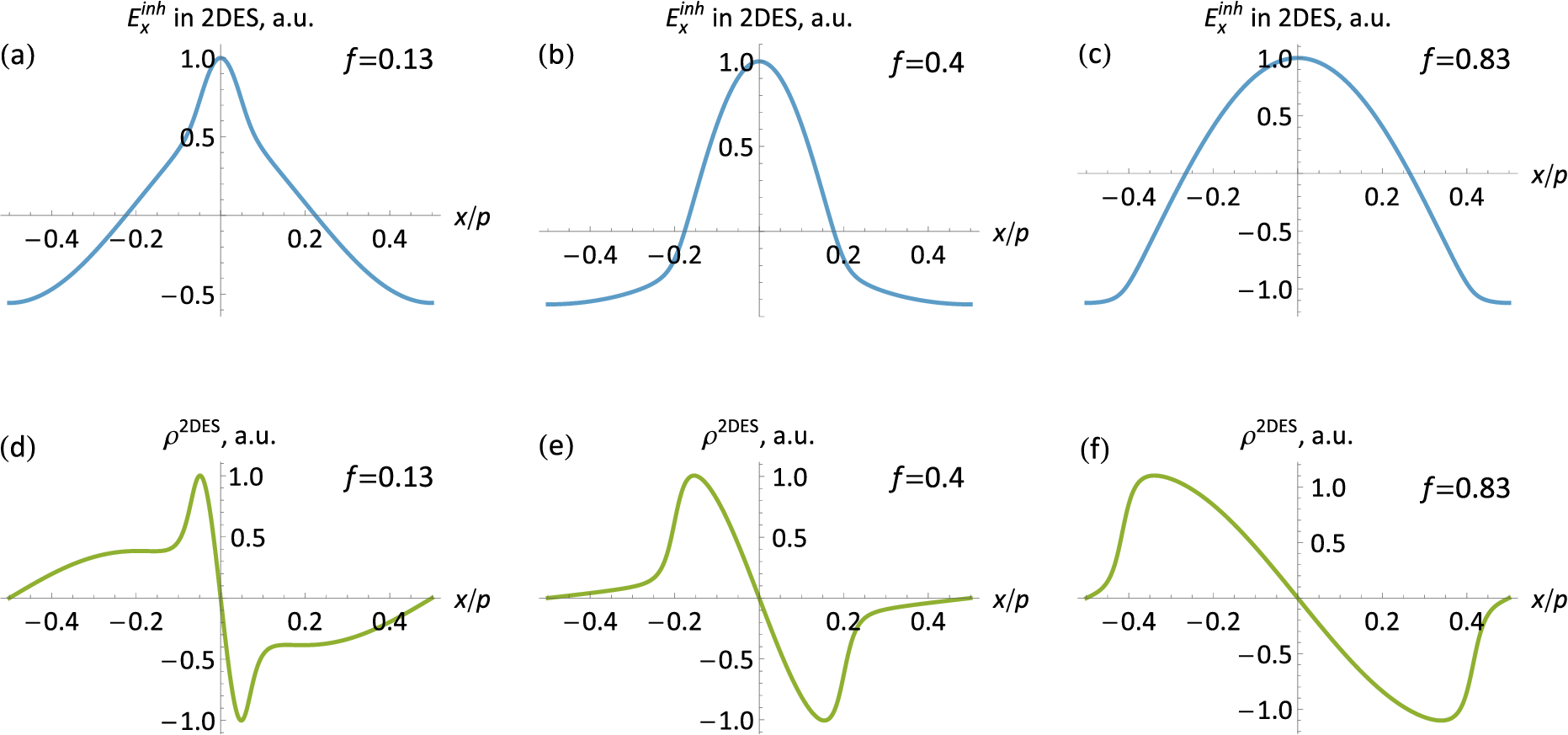}
    \caption{Panels (a)--(c) show the inhomogeneous part of $x$-component of electric field in 2DES, $E_x^{inh}$, as a function of coordinate $x$ for different filling factors $f=l_g/p$: $0.13$, $0.4$, and $0.83$ (here $l_g$ is the width of the gate and $p$ is the period of the structure). Panels (d)--(f) show the charge density in 2DES, $\rho^{2DES}$, for the same filling factors. The gated part of 2DES extends from $-f/2$ to $f/2$ in dimensionless coordinates $x/p$.
    }
    \label{S4}
\end{figure}

The inhomogeneous part of electric field, $E_x^{inh}(x)$, found from Eq.~(\ref{Supp:ExInh}), divided by the value $E_x^{inh}(x=0)$, is presented in Fig.~\ref{S4}(a)--(c); hereinafter we left 100 terms in the sum by $k$ in our calculations.

Finally, we analyze the charge density in 2DES, $\rho^{2DES}(x)$. As the 2DES is homogeneous, i.e. its conductivity is constant, we can use the following ordinary relations
\begin{equation}
\label{Supp:ro}
    \rho^{2DES}(x) \propto \text{div}\, {\bf j}^{2DES} \propto \frac{\partial E_x^{inh}}{\partial x} \propto -\sum_{k=1}^{+\infty} 2 \pi k \,F(k) \sin 2 \pi k x.
\end{equation}
Note that the homogeneous part of the density current and $x$-component of the electric field do not contribute to the charge density. The last relation in Eq.~(\ref{Supp:ro}) allows us to find the charge density in 2DES, which (in arbitary units) is presented in Fig.~\ref{S4}(d)--(f).